# Pore-scale insights into the role of micro fractures on permeability of fractured porous media

Ruichang Guo, Hongsheng Wang, Reza Ershadnia, Seyyed Hosseini*

Bureau of Economic Geology, Jackson School of Geosciences, The University of Texas at Austin, Austin, TX 78758, USA

**Abstract**

Fractures play a critical role in governing fluid flow within subsurface energy systems, including oil and gas production, geologic carbon sequestration, and underground hydrogen storage. This study investigated the impact of pore-scale fractures on fluid flow and permeability in fractured porous media. The analysis focused on a single fracture embedded within a porous medium. Fluid flow was simulated using the lattice Boltzmann method, and the effects of fracture length, width, and orientation angle on permeability were systematically examined. Results showed that increasing both fracture length and width enhanced permeability. Additionally, fractures oriented more closely to the flow direction (i.e., smaller orientation angles) resulted in higher permeability. Interestingly, when the orientation angle approached 90°, the presence of a fracture could reduce the overall permeability of the porous medium. A critical orientation angle was identified, beyond which the fracture decreased permeability; this critical angle was found to increase with fracture width. Permeability tensors were also fitted to determine the critical angle and quantify the influence of fracture width on the critical orientation angle. These findings provide new insights into the role of microfractures in controlling permeability, with important implications for subsurface energy systems.

**Key words**： fluid flow, fractured porous media, permeability tensor, subsurface energy system

*Corresponding author email: seyyed.hosseini@beg.utexas.edu

## 1. Introduction

Fluid flow in porous media plays a critical role in a wide range of subsurface energy systems (National Academies of Sciences, 2015; Dejam et al., 2018; Zamehrian et al., 2022; Yu et al., 2024; Guo et al., 2025b). Fractures are commonly present in formation rocks and can arise from both natural geologic processes and engineered interventions (Kranz, 1983; Anders et al., 2014; Liu et al., 2021; Qu et al., 2023). Permeability is a key transport property governing fluid movement in fractured porous media (Pirker et al., 2008; Delle Piane et al., 2015; Phillips et al., 2020; Liu et al., 2022). Although fractures typically occupy a small volume fraction of the rock, their influence on overall permeability can be substantial (Pakdel and Pietruszczak, 2024). Accurately characterizing the equivalent permeability of fractured formations is essential for evaluating the performance of subsurface energy systems and has remained an active area of research for decades (Pickup et al., 1994; Olson, 2007; Jafari and Babadagli, 2012; Chen and Jiang, 2015; Liang et al., 2023).

A fracture in a geological porous material refers to a discontinuity or narrow zone with significantly higher permeability than the surrounding matrix, often serving as a preferential pathway for fluid flow and solute transport. In simplified representations of a fracture as a space between two plates, fracture permeability is described by the Cubic Law, which relates permeability solely to the fracture aperture (Zimmerman and Bodvarsson, 1996). However, natural fracture geometries are far more complex. Fluid flow within fractures is influenced not only by aperture but also by surface roughness and contact morphology (Renshaw and Park, 1997; Zhang and Chai, 2020; Phillips et al., 2021; Li et al., 2025). Beyond individual fracture properties, the equivalent permeability of fractured porous rocks is strongly affected by the orientation, spacing, and connectivity of the fracture network (Pakdel and Pietruszczak, 2024). While fractures can significantly increase the absolute permeability of a porous medium (Delle Piane et al., 2015; Li et al., 2025), their role becomes even more critical in systems with fully developed fracture networks, where fractures act as the primary flow conduits, and the porous matrix serves as the main fluid storage domain. Effective fluid exchange between fractures and the matrix is essential to the performance of many subsurface engineering processes. In reservoirs, fractures typically exist as interconnected networks. The inherent randomness of fracture distribution and the geometric complexity of fracture networks make the modeling of fractured porous media a considerable challenge.

There are generally three main approaches for simulating fluid flow in fractured reservoirs (Lee et al., 1999; Berre et al., 2019). The first is the single-continuum approach, in which the effects of fractures are incorporated into an effective or equivalent permeability tensor for the porous medium (Lian and Cheng, 2012; Fahad et al., 2017; Berre et al., 2019). The second is the dual-continuum approach, including models such as the dual-porosity and dual-permeability models (Barenblatt et al., 1960). In this framework, the fractured porous medium is represented by two overlapping continua: a high-permeability continuum for the fracture network and a high-storage continuum for the porous matrix (Izadi et al., 2009). The third approach is the explicit representation of fracture networks, which includes the discrete fracture network (DFN) and discrete fracture–matrix (DFM) models. In these models, fractures are treated as high-permeability features with their geometry explicitly resolved (Bogdanov et al., 2003; Fu and Yang, 2022). The key distinction lies in the treatment of the matrix: the DFN model assumes an impermeable matrix and considers flow only within the fractures (Lee et al., 1999; Teimoori et al., 2003), whereas the DFM model treats the matrix as a permeable domain and accounts for fluid flow in both the fractures and the surrounding matrix (Berre et al., 2019).

The selection and application of these modeling approaches must be consistent with the scale of the fracture network. When the fracture network exhibits a broad range of length scales, small-scale fractures which are smaller than the grid block size of the reservoir model still need to be homogenized, even within the DFM framework, by computing an equivalent permeability. Given the inherent complexity of fractured porous media, the permeability tensor is commonly employed to represent the equivalent permeability of the fractured system. Several methods have been proposed to estimate this tensor, accounting for fracture geometry and connectivity (Oda, 1985; Bourbiaux et al., 1998; Min et al., 2004).

Previous studies have primarily focused on either large-scale fracture networks or the upscaling of transport properties in fractured porous media. However, natural reservoirs often contain a large number of microfractures (Laubach et al., 2004; Laubach and Diaz-Tushman, 2009; Anders et al., 2014). Examples of rock samples containing microfractures are shown in **Fig. 1**. A fracture can either enhance or reduce the permeability of a porous medium: when filled with fine-grained materials, it typically reduces the equivalent permeability, while an empty fracture is generally assumed to increase it. However, our recent findings suggest that microfractures do not necessarily enhance permeability. The influence of a single microfracture on the effective permeability remains poorly understood and warrants further investigation.

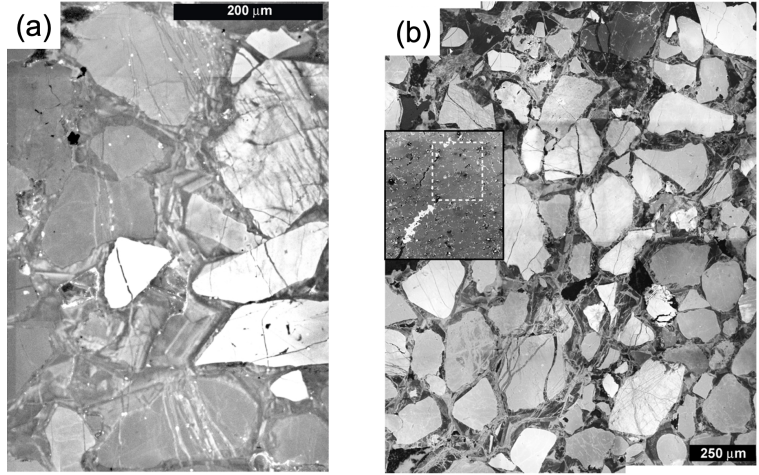

Fig. 1. Examples of microfractured rock samples (Laubach et al., 2004). (a) Quartz-cemented quartzarenite from the Cretaceous Travis Peak Formation, East Texas, USA. (b) Quartz-cemented quartzarenite from the Cretaceous Fall River Formation, Powder River Basin, Wyoming, USA.

When the length scale of fractures is comparable to the pore size, the difference in fluid flow between the fracture and the porous matrix is primarily governed by the contrast in their pore structures. To uncover the mechanisms by which microfractures influence fluid flow in porous media, pore-scale investigations are essential. With the rapid advancement of experimental techniques (Zhao et al., 2016; Guo et al., 2025a) and pore-scale simulation methods (Liang et al., 2023; Li et al., 2024; Zhao et al., 2024), such studies have significantly deepened our fundamental understanding of fluid transport in complex porous systems. In this work, pore-scale direct simulations using the lattice Boltzmann method were employed to investigate the effects of key fracture parameters on fluid flow and the resulting permeability in fractured porous media.

This study aimed to address the knowledge gap regarding the role of microfractures in fluid flow through fractured porous media. The investigation focused on a single microfracture whose width was comparable to the average pore size, and its effect on the overall permeability of the medium. Pore-scale direct simulations were conducted to model fluid flow in the fractured porous domain. The methodology and construction of the fractured porous media model were presented in Section 2. Section 3 presented the results on flow field characteristics and analyzes how fracture parameters influence permeability. In Section 4, a critical orientation angle was proposed, and its determination as well as its dependence on fracture width were discussed. Concluding remarks were provided in Section 5.

## 2. Materials and methodology

### 2.1. Fractured porous media

The homogeneous porous medium is shown in **Fig. 2**, with dimensions of 6 mm × 6 mm. The porosity is 40.78%, and the average pore size is 72.89 μm. A simplified rectangular fracture is embedded at the center of the domain. The fracture is characterized by its length ($L$), width ($W$), and orientation angle ($\theta$). Its center is aligned with the center of the porous medium. The $\theta$ is defined as the angle between the fracture and the fluid flow direction and is varied from 0° to 90°. The fracture length is normalized by the side length of the porous medium, and the fracture width is normalized by the average pore size. The detailed parameters of the fractured porous medium are provided in **Table 1**.

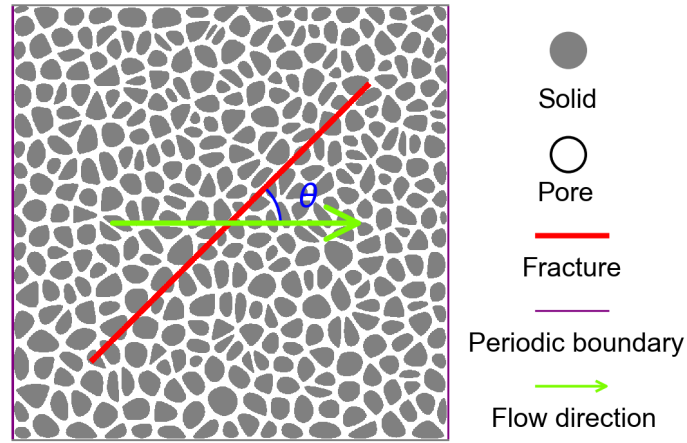

Fig. 2. Pore structure of the fractured porous medium.

Table 1. Fracture parameter setting of the fractured porous medium

| | Physical | | Normalized |
|---|---|---|---|
| $L$, μm | 600, 3000, 5400 | $L^*$ | 0.1, 0.5, 0.9 |
| $W$, μm | 36.44, 72.89, 109.34, 145.78 | $W^*$ | 0.5, 1.0, 1.5, 2.0 |
| $\theta$, ° | 0, 30, 45, 60, 90 | | |

### 2.2. Numerical simulation

Pore-scale direct simulation offers significant advantages in capturing microscopic fluid flow processes within porous media. In this study, direct numerical simulations based on the lattice Boltzmann method were employed to model fluid flow in fractured porous media. The lattice Boltzmann method serves as an alternative numerical approach to solving the Navier–Stokes equations and is particularly well-suited for complex geometries. Further details on the verification and implementation of the lattice Boltzmann method can be found in Guo et al. (2025c).

The resolution of the domain was 6.67 µm/pixel. The fluid in the simulation was water. During the simulations, periodic boundary conditions were applied at the inlet and outlet, as illustrated in Fig. 2, while solid surfaces were treated as no-slip wall boundaries. A body force was applied to drive the flow through the domain. The relative permeability was then calculated using Darcy's law.

## 3. Results
### 3.1. Flow fields of fractured porous media

It was well established that fractures in porous media generally act as high-conductivity channels, enhancing the equivalent permeability of fractured porous media. However, when fracture dimensions were comparable to the pore scale, their impact became highly dependent on both geometry and orientation. Specifically, when $W^*$ was smaller than 1 or $\theta$ approached 90°, the fracture may fail to establish an effective high-conductivity pathway. Representative flow fields for selected fractured porous media were shown in **Fig. 3**, illustrating the influence of fracture geometry and orientation on fluid flow patterns within the domain.

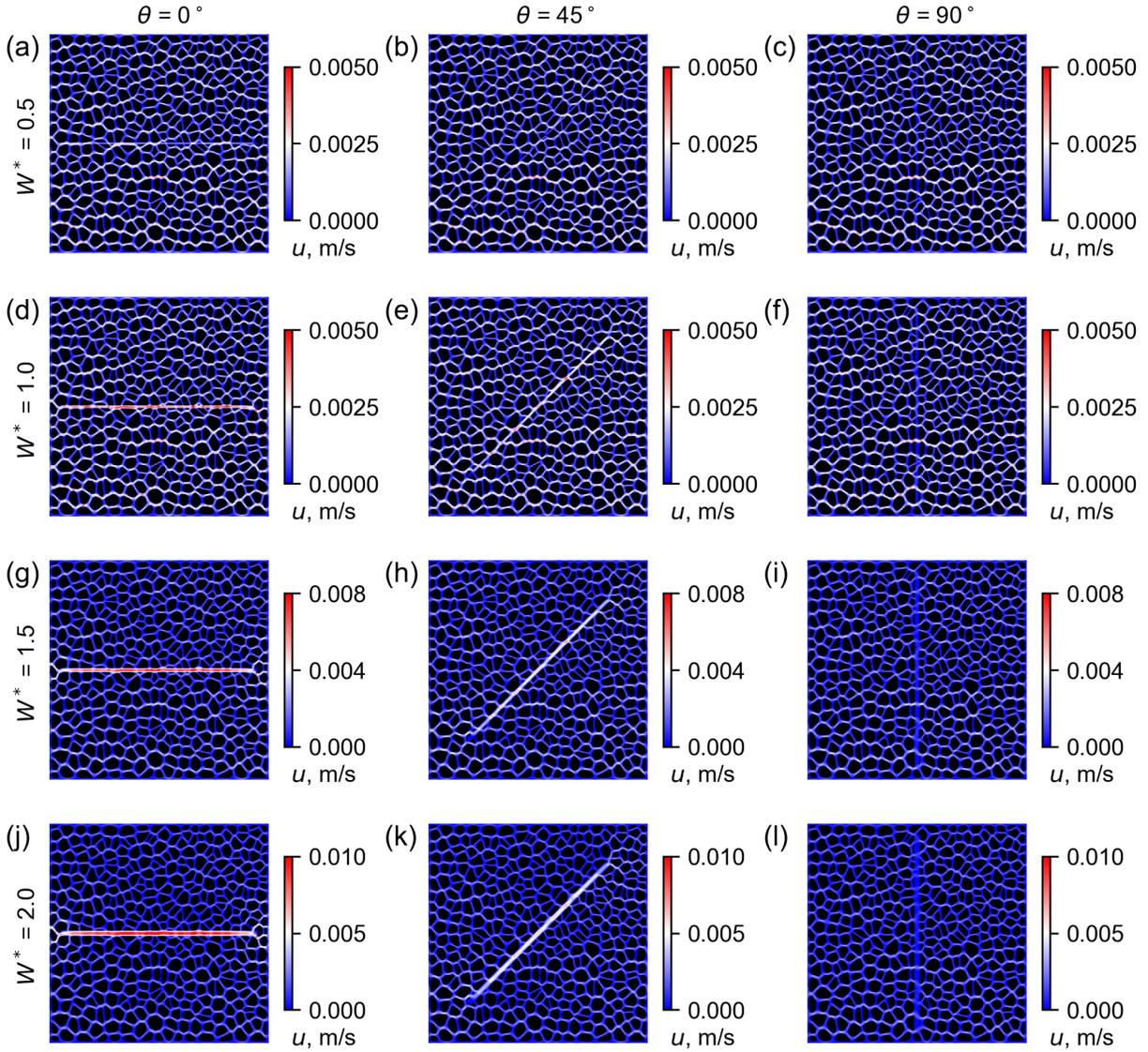

**Fig. 3.** Flow fields in fractured porous media with $L^* = 0.9$. Each row corresponds to a different: $W^* = 0.5$, 1.0, 1.5, and 2.0, respectively. Each column represents a different $\theta = 0°$, 45°, and 90°, respectively. Solid grains within the porous medium are shown in black.

The influence of a microfracture on fluid flow depended strongly on the $W^*$. When $W^* < 1.0$, the fluid velocity within the fracture was not significantly higher than that in the surrounding pore space. As illustrated in Figs. 3a–c ($W^* = 0.5$), the velocity within the fracture was only marginally greater than in the matrix. In contrast, larger values of $W^*$ promoted the development of higher flow velocities within the fracture. As shown in Fig. 3, fractures with greater $W^*$ exhibited clearly enhanced velocity compared to the surrounding porous medium.

A microfracture enhanced fluid velocity by reducing local tortuosity; however, this effect was strongly dependent on the $\theta$. When $\theta = 0°$, the fracture was aligned with the flow direction, minimizing the local tortuosity and forming a high-velocity channel. As $\theta$ increased, the ability of the fracture to reduce tortuosity and facilitate flow gradually diminished. At $\theta = 90°$, the fracture was perpendicular to the flow direction, and the local tortuosity became comparable to that of the surrounding matrix. Consequently, the velocity within the fracture was only slightly higher than in the adjacent pore space.

To further illustrate the effect of fractures on fluid velocity, the average velocity $\bar{u}$ within the pore space (excluding the fracture) and within the fracture itself was plotted in **Fig. 4**. The results showed that when the fracture was aligned with the flow direction, fluid velocity within the fracture developed significantly. For example, when $W^* = 1$, the average velocity within the fracture was approximately 3.1 times greater than that in the surrounding pore space. This enhancement increased with fracture width: at $W^* = 2$, the velocity within the fracture was 7.0 times higher than that in the pore space.

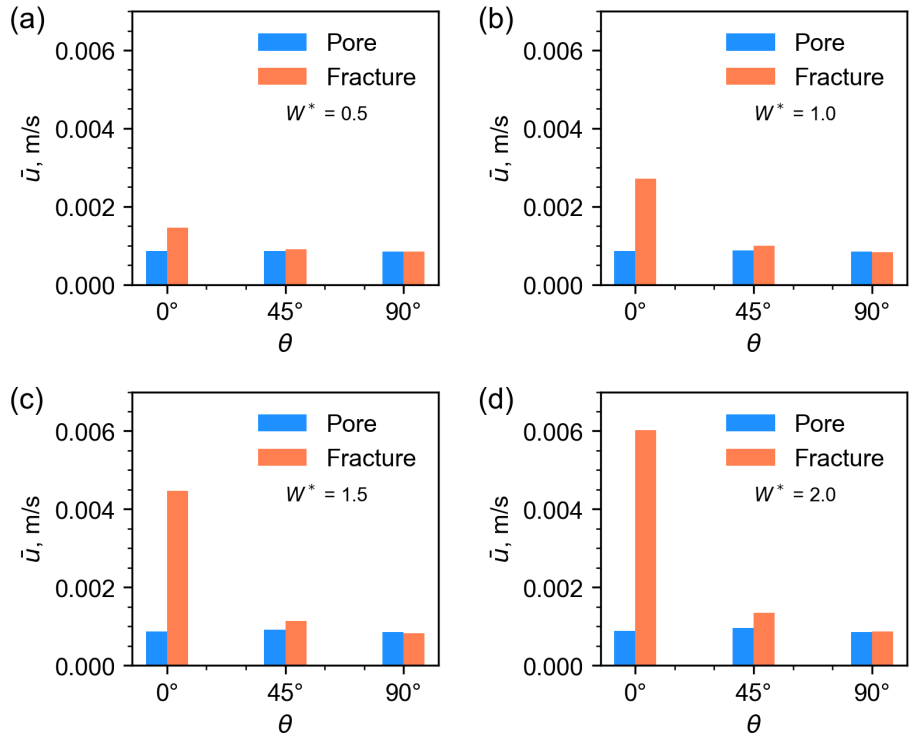

**Fig. 4** Average velocity in the pore space and in the fracture for (a) $W^* = 0.5$, (b) $W^* = 1.0$, (c) $W^* = 1.5$, (d) $W^* = 2.0$, respectively.

## 3.2. Impact on permeability

Permeability is a fundamental parameter that quantifies the ability of a porous medium to transmit fluids. To better assess the impact of fractures, the normalized permeability of the fractured porous medium was used, defined as $k^* = k/k_0$, where $k_0$ was the permeability of the medium with no fracture.

Figure 5a showed the relationship between $k^*$ and $\theta$ for different $L^*$. Consistent with the trends observed in the velocity fields (Fig. 3), increasing $\theta$ led to a reduction in $k^*$, with the decline being more pronounced for larger $L^*$. This indicated that longer fractures more effectively promoted velocity development and the formation of high-conductivity flow channels, thereby enhancing permeability. Interestingly, the three $k^*$-$\theta$ curves intersected at a common point where $k^* = 1$, suggesting the existence of a critical orientation angle, denoted as $\theta_{cr}$. When $\theta$ was larger than the $\theta_{cr}$, the $k^*$ fells below 1 regardless of fracture length, indicating that the presence of the fracture reduced permeability in all cases.

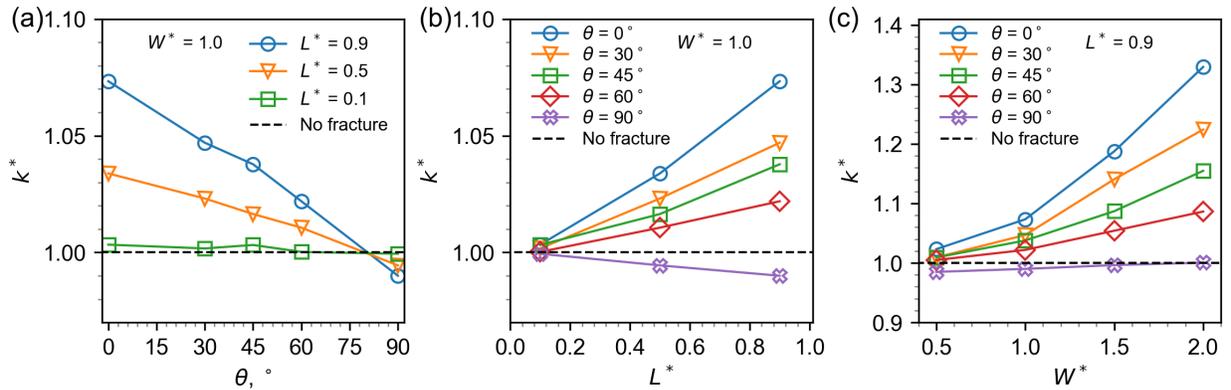

**Fig. 5** Impact of (a) $\theta$, (b) $L^*$, and (c) $W^*$ of microfractures on $k^*$ of the fractured porous media.

The relationship between $k^*$ and $L^*$ was shown in Fig. 5b. As observed, increasing fracture length enhances permeability when $\theta = 60°$. However, for $\theta = 90°$, longer fractures led to a reduction in $k^*$. This suggested that when $\theta$ exceeded the critical angle $\theta_{cr}$, fractures consistently reduced permeability, and the magnitude of this reduction increased with fracture length. For a fixed $\theta$, the influence of a microfracture appeared to accumulate linearly with increasing $L^*$, indicating a scaling relationship of the form $k^* \propto L^*$.

The effect of $W^*$ on $k^*$ was shown in Fig. 5c. As observed, increasing $W^*$ enhanced $k^*$ across all orientation angles $\theta$, including cases where $\theta > \theta_{cr}$. For an ideal two-dimensional fracture with width $b$, the theoretical scaling law is $k \propto b^2$ (Zimmerman and Bodvarsson, 1996). Based on this, it could be inferred that $k^*$ followed a power-law relationship with $W^*$, expressed as $k^* \propto W^a$. When $\theta = 0°$, the exponent $a$ approached 2, consistent with the ideal case. In contrast, for $\theta = 90°$, the exponent $a$ approached 1. For intermediate orientations (0° < $\theta$ < 90°), the exponent $a$ lied between 1 and 2.

To further examine the existence of the critical orientation angle $\theta_{cr}$, $k^*$- $\theta$ relationships for different values of normalized fracture width $W^*$ were plotted in **Fig. 6**. The trends were consistent with those observed for $W^* = 1.0$ in Fig. 5a, showing that the influence of $\theta$ became more pronounced as $W^*$ increased. As illustrated in Fig. 6, for $W^* = 0.5$ and $W^* = 1.5$, a critical angle $\theta_{cr}$ existed, beyond which $k^*$ fell below 1—indicating that the presence of the fracture reduced permeability. However, when $W^* = 2.0$, all values of $k^*$ remained greater than 1 across the full range of $\theta$, suggesting that a critical orientation angle did not exist for sufficiently large fracture widths.

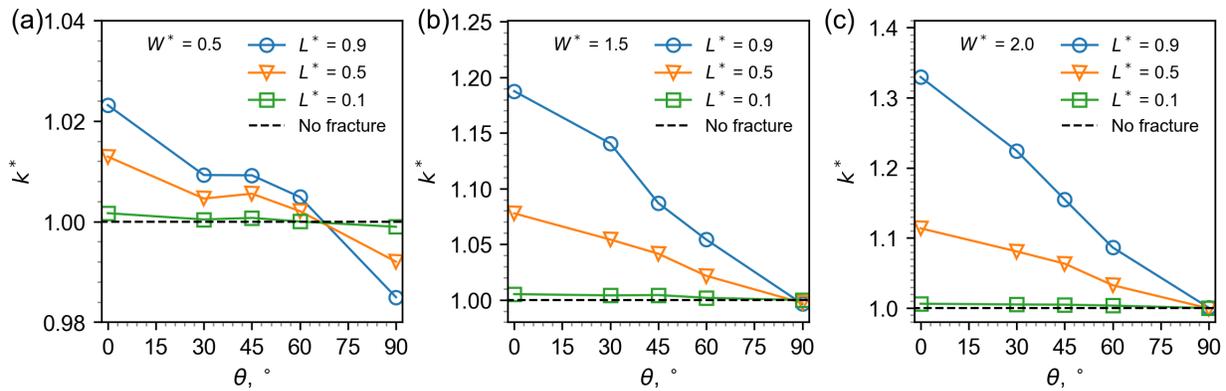

**Fig. 6** Impact of microfracture orientation angle on the permeability of the fractured porous medium with $W^*$ of (a) 0.5, (b) 1.5, (c) 2.0.

## 4. Discussion

In this section, the determination of the critical orientation angle $\theta_{cr}$ and its sensitivity to fracture width were discussed. A common representation of permeability in porous media is through a permeability tensor, which captures the directional dependence of flow properties (Scheidegger, 1954; Oda, 1985; Gupta et al., 2001; Bear, 2013). The unfractured porous medium considered in this study was homogeneous and isotropic. For the fractured porous media, the permeability tensor was symmetric and expressed as (Pickup et al., 1994):

$$\mathbf{k}^* = \begin{bmatrix} k_x^* \cos^2\theta + k_y^* \sin^2\theta & (k_x^* - k_y^*)\sin\theta\cos\theta \\ (k_x^* - k_y^*)\sin\theta\cos\theta & k_x^* \sin^2\theta + k_y^* \cos^2\theta \end{bmatrix} \tag{1}$$

where $k_x^*$ is the normalized permeability in the $x$ direction, i.e. $\theta = 0°$; $k_y^*$ is the normalized permeability in the $y$ direction, i.e. $\theta = 90°$.

The permeability data were used to fit the permeability tensors of the fractured porous media, and the resulting tensors were shown in Fig. 7. For a given fracture width ($W^* = 0.5$, 1.0, or 1.5), all fitted permeability tensors intersected with that of the unfractured medium at a single point, regardless of $L^*$. The corresponding orientation angle at this intersection was defined as the critical angle $\theta_{cr}$. This value was determined by averaging the intersection angles between the permeability tensors of fractured media with varying $L^*$ and the tensor of the unfractured medium. When $W^* = 2.0$, all permeability values exceeded that of the unfractured medium for the full range of $\theta$, indicating that $\theta_{cr}$ did not exist for sufficiently large fracture widths.

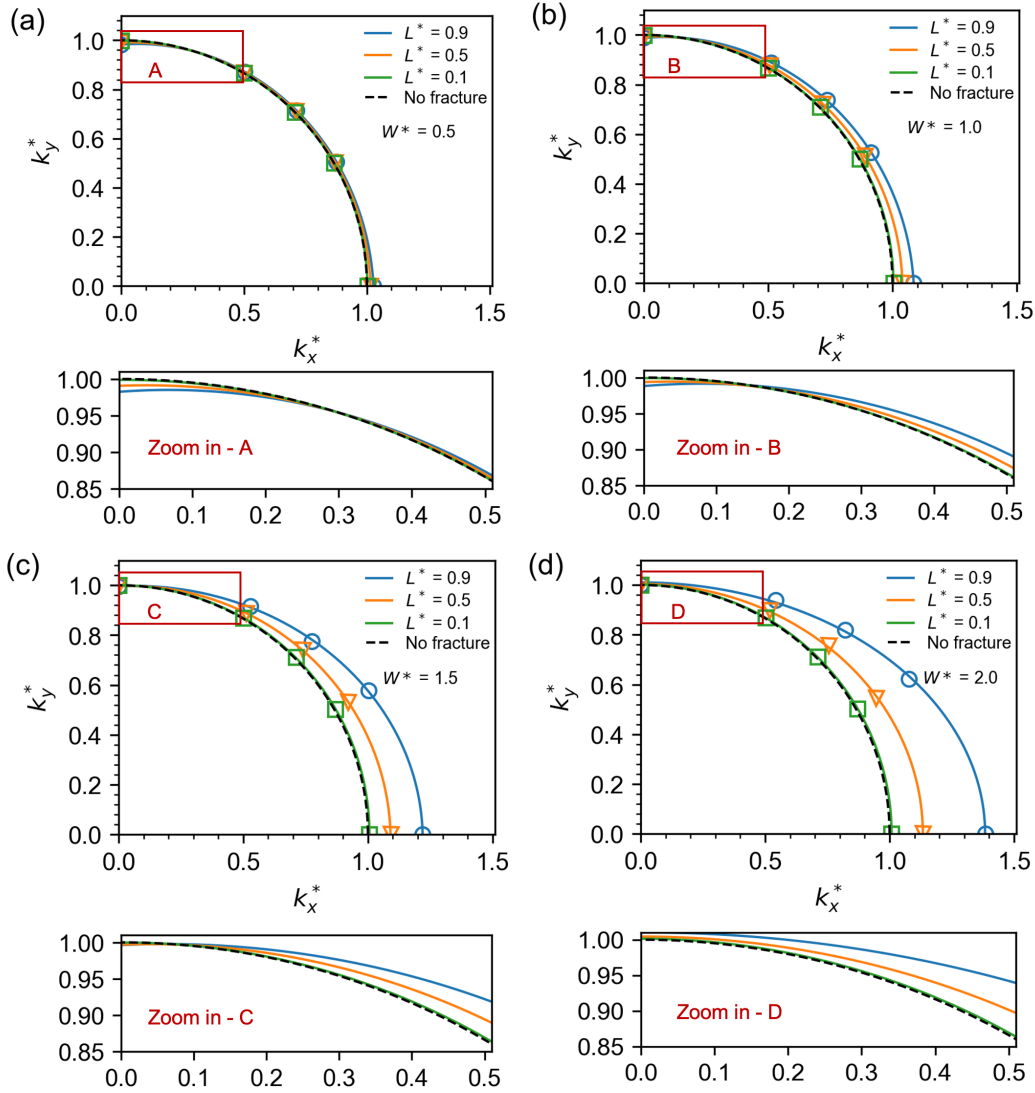

**Fig. 7**. Impact of $\theta$ of microfractures on $k$ of the fractured porous medium with different $W^*$. The solid lines were fitted curve. The dotted lines were the permeability of the porous medium with on fracture. The data points were measurements with numerically simulated results.

The relationship between the $\theta_{cr}$ and $W^*$ was shown in Fig. 8. A distinct $\theta_{cr}$ was observed when the fracture width was comparable to the average pore size. However, when $W^* = 2.0$, the permeability of all fractured porous media exceeded that of the nonfractured medium across all orientation angles. In this case, $\theta_{cr}$ either exceeded 90° or ceased to exist.

This finding suggested that a fracture could reduce the permeability of a porous medium when the orientation angle $\theta$ exceeds the critical angle $\theta_{cr}$, challenging the common assumption that fractures always enhanced permeability.

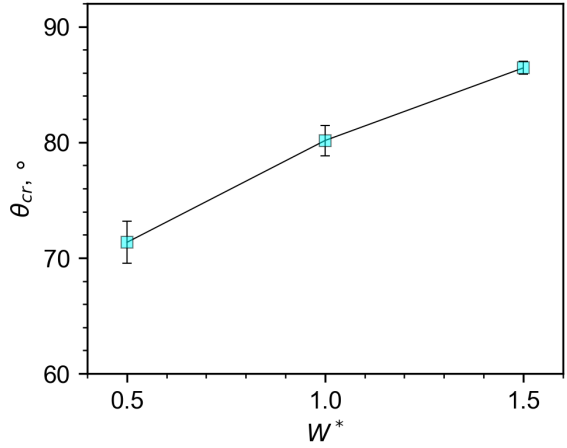

**Fig. 8.** $\theta_{cr}$- $W^*$ relation.

## 5. Conclusions

The impact of a microfracture on the permeability of a homogeneous porous medium was investigated using pore-scale direct simulations based on the lattice Boltzmann method. In this study, the fracture width $W^*$ ranged from 0.5 to 2.0 times the average pore size. The effects of fracture length $L$, width $W$, and orientation angle $\theta$ on the permeability of fractured porous media were systematically examined. The results revealed that under certain conditions, a microfracture can reduce the permeability of the medium. A critical orientation angle $\theta_{cr}$ was identified, which delineated whether a fracture enhanced or diminished permeability. The main conclusions were summarized as follows:

(1) A fracture did not always increase the permeability of a porous medium; its effect depended on the $\theta$. A critical angle $\theta_{cr}$ was identified, such that when $\theta > \theta_{cr}$, the fracture consistently reduced permeability. When $\theta < \theta_{cr}$, the fracture enhanced permeability.

(2) The influence of $L$ on permeability depended on $\theta$. For $\theta > \theta_{cr}$, increasing $L$ led to higher permeability, and the effect became more significant as $\theta$ approached 0°. Conversely, when $\theta < \theta_{cr}$, longer fractures reduced permeability.

(3) Increasing $W$ consistently increased permeability, regardless of $\theta$. However, even at larger $W$, when $\theta > \theta_{cr}$, the permeability remained lower than that of the homogeneous, nonfractured porous medium.

(4) The value of $\theta_{cr}$ was controlled by $W$. As $W$ increased, $\theta_{cr}$ also increased and eventually disappeared upon reaching 90°. In the porous medium studied here, $\theta_{cr}$ ceased to exist when the fracture width approached twice the average pore size.

This study is the first to demonstrate that pore-scale fractures can reduce the permeability of a porous medium and introduces the concept of a critical orientation angle $\theta_{cr}$ to identify the conditions under which a fracture diminishes permeability. This finding advances our understanding of the role of microfractures in controlling fluid flow in porous media.

**Acknowledgement**

The authors acknowledge the financial support from the GeoH$_2$ Industrial Affiliates Program at the University of Texas at Austin. Preparation of this manuscript was partly supported by a Publication Grant from the Bureau of Economic Geology